\pdfoutput=1

\documentclass[11pt]{article}

\usepackage[final]{acl}
\usepackage{multirow} 
\usepackage{listings} 
\usepackage{times}
\usepackage{latexsym}
\usepackage{amsmath}

\usepackage[T1]{fontenc}

\usepackage[utf8]{inputenc}

\usepackage{microtype}

\usepackage{inconsolata}

\usepackage{graphicx}
\usepackage{tablefootnote}

\begin{document}
\title{Just Put a Human in the Loop? \\ Investigating LLM-Assisted Annotation for Subjective Tasks}


\author{
Hope Schroeder \quad Deb Roy \quad Jad Kabbara \\
Massachusetts Institute of Technology \\
\texttt{\{hopes, dkroy, jkabbara\}@mit.edu}
}

\maketitle

\begin{abstract}
LLM use in annotation is becoming widespread, and given LLMs' overall promising performance and speed, simply "reviewing" LLM annotations in interpretive tasks can be tempting. In subjective annotation tasks with multiple plausible answers, reviewing LLM outputs can change the label distribution, impacting both the evaluation of LLM performance, and analysis using these labels in a social science task downstream. 
We conducted a pre-registered experiment with 410 unique annotators and over 7,000 annotations testing three AI assistance conditions against controls, using two models, and two datasets. We find that presenting crowdworkers with LLM-generated annotation suggestions did not make them faster, but did improve their self-reported confidence in the task. More importantly, annotators strongly took the LLM suggestions, significantly changing the label distribution compared to the baseline. 
When these labels created with LLM assistance are used to evaluate LLM performance, reported model performance significantly increases. 
We believe our work underlines the importance of understanding the impact of LLM-assisted annotation on subjective, qualitative tasks, on the creation of gold data for training and testing, and on the evaluation of NLP systems on subjective tasks.
\end{abstract}

\section{Introduction}

Large language models (LLMs) have shown impressive performance in many annotation tasks, including subjective tasks common in content moderation and text analysis in the social sciences. 
Evaluating human annotation of subjective tasks for comparison against LLM annotation performance, either for the task of end-to-end qualitative analysis or for the construction of ground truth for NLP tasks, is difficult in the absence of domain experts. Accordingly, hiring a large number of crowd annotators (often in service of creating a crowd decision) becomes attractive in the evaluation of NLP systems on social science tasks. However, managing and paying crowdworkers can be difficult, and crowdworkers often have varied performance.

A significant body of research explores how AI suggestions can assist qualitative researchers \cite{jiang_supporting_2021, feuston_putting_2021, overney_sensemate_2024}. 
Labeling text according to a complex qualitative "codebook" is a repetitive, time-consuming task, and advances in LLM capabilities have made using LLMs in annotation attractive \cite{wang_want_2021_fixed}. Complex, theory-driven text analysis is increasingly being mediated by LLMs \cite{de_paoli_can_2023} or LLM-powered tools \cite{lam_concept_2024}.


Given all this progress, LLMs have created opportunities to create annotation pipelines that appear to work off the shelf without fine-tuning, making automated annotation accessible to practitioners with less technical skill. LLMs' reported performance in annotating socially complex topics \cite{gilardi_chatgpt_2023}, sometimes with greater skill than humans \cite{he_if_2024}, potentially opens LLM-based annotation to an even wider range of fields and practices compared to past years. This makes understanding the many ways humans may interact with LLM annotations more important.

LLMs increasingly power both software interfaces and complex research tools \cite{pang_understanding_2025}. Generating plausible LLM annotations for a variety of tasks is now easy, enabling potentially powerful analysis. Some systems suggest that putting a "human in the loop" to check annotations can ensure the model’s outputs are "reasonable and
reliable" \cite{wang_media_2025}. 
But given that we know humans are subject to anchoring bias---the bias towards the first option we are presented, especially if it is in the plausible range \cite{doi:10.1126/science.185.4157.1124}---humans may review and confirm LLM annotations that are plausible, but nonetheless significantly change 1) the annotation evaluation process and 2) the outcome the annotations get used for (such as the decision boundary for classification judgments, or the distribution of annotations used in a text analysis), downstream. 

In this work, we ask several questions. First, when provided with (different forms of) LLM assistance, do crowdworker annotators "produce more" by going faster in a complex, subjective annotation task, and does this result in them "understanding less?" Do they find LLM suggestions accurate and helpful, and how often do annotators take these suggestions? If annotations influenced by LLM suggestions are used to construct ground truth on annotation tasks involving complex social concepts, what effect does that have on evaluated performance of these LLMs on those annotation tasks? 

We use two community conversation datasets to address these questions, recruiting crowdworkers to annotate the data according to a list of identified themes. These workers are presented with LLM-suggested annotations in a variety of formats. We then study how the crowdworkers use these LLM suggestions in a complex annotation process in service of answering these research questions.

Our findings open up questions regarding the use of AI assistance in qualitative research where annotators act as independent reviewers of AI suggestions and can supposedly still retain full control of the analysis. 
We believe our work underlines the importance of understanding the impact of LLM-assisted annotation on subjective, qualitative tasks and the interfaces that mediate them, on the creation of gold data for training and testing, and on the evaluation of NLP systems on subjective tasks. In addition to the findings that answer the aforementioned questions, we release a dataset of human and LLM-assisted annotations on two complex qualitative codebooks across a variety of conditions.

\section{Related Work}

LLMs have been increasingly employed in various "human in the loop" AI assistance setups, e.g., in education \cite{jiang-etal-2024-leveraging} and coding \cite{mozannar2024show}. LLMs have also been used for annotation tasks, for reasons like significantly decreasing labeling cost \cite{wang_want_2021_fixed}.
\citet{ziems_can_2023} discussed how LLMs are being widely used in computational social science tasks, including subjective tasks, often with success. 
\citet{li_synthetic_2023} found that the performance of LLMs trained on synthetically generated ground truth data is negatively associated with the subjectivity of the task. More work has shown the challenges of LLM annotation: LLM annotation performance can be highly variable to prompts \cite{atreja_prompt_2024} and can depend on the ordering of choices presented in a prompt \cite{wang_primacy_2023, liu2023lostmiddlelanguagemodels}.

While using LLMs to create annotations may have "good enough" performance and likely does decrease cost compared to hiring human crowd workers, many tasks still require or benefit from human involvement "in the loop," especially for subjective annotation and analysis tasks; some methods suggest ways of using humans in the loop to optimize prompts for LLM annotation \cite{pangakis2024keeping}. Many AI-assisted text annotation platforms have also been developed and published within the HCI space \cite{overney_sensemate_2024, gao_collabcoder_2024, gao_coaicoder_2023, gao_using_2025}, accelerating this change. Research shows AI-suggested labeling platforms increase agreement and convergence on qualitative codes, or text labels \cite{10.1145/3617362}.

There is increasing evidence that humans tend to anchor on the suggestions that AI systems give, which can result in changes to communication and even user opinions \cite{jakesch}. Some work observed how humans anchor on LLM outputs in text analysis tasks, including topic induction. For instance, \citet{choi2024llmeffecthumanstruly} show that in a topic generation task, analysts anchor on LLM outputs, resulting in different topic lists depending on whether or not they saw the LLM versions.
This illustrates the potential risk of homogenization of insight as a result of AI influence on text analysis. This concern is raised by \citet{messeri_artificial_2024} regarding AI's influence on science more generally, and the authors discuss LLMs' potential to reduce diversity in human judgment, creating an environment in science where we "produce more but understand less."

Many cite time savings as their motivation for using LLMs as annotators, however, findings have so far been mixed regarding 
productivity. For example, \citet{Bughin27112024} shows that while AI can boost coding productivity, there exists a tradeoff between productivity and coding quality, and \citet{overney_sensemate_2024} observe that users with AI support in the SenseMate platform spent more time on qualitatively coding data.

Here, we examine how presenting LLM-generated suggestions to annotators in a complex subjective annotation task affects their self-reported understanding of the task as well as the overlap of crowd decisions on text labels with LLM annotations of the same labels, with implications for the evaluation of LLM performance on these tasks, even when humans are put "in the loop" to review and confirm annotations.

\section{Data and Codebook}

We source data from the Fora corpus \cite{schroeder_fora_2024}. In 2022, the NYC Department of Health \& Mental Hygiene, the NYC Public Health Corps (PHC), and the non-profit \href{https://cortico.ai/}{Cortico} recruited over 100 communities to a series of 28 small group dialogues hosted in New York City to understand community resourcing and vaccine decisions during COVID-19. Following the conversations, community workers created a codebook of themes of interest to the "NYC" corpus (as we will call it), then labeled quotes from the conversations with the themes denoted by the codebook. Similarly, in 2021, a conversation series in Boston called "Real Talk for Change" was hosted to understand issues in marginalized communities leading up to the 2021 Boston Mayoral election. We use the conversation data and codebook created for this "RTFC" corpus as well. 


The NYC codebook developed by community partners had 7 overarching themes related to health and vaccine decisions, including \textit{External Motivations: Friends \& Family}, \textit{Intrinsic Motivations: Not wanting to get the virus},  and \textit{Role of Community Health Organizations: Health Education \& Support}, and \textit{Vaccine Hesitancy}. Each of the 7 top-level labels had sublabels, for a total of 20 unique labels related to the NYC corpus identifying phenomena of interest to the community partner. Similarly, the RTFC corpus has 9 overarching top-level labels, and 41 sublabels relevant to the corpus, such as \textit{Safety: Street violence} and \textit{Housing: Housing affordability}. The full codebook for each corpus is available in the appendix. We sampled 200 quotes\footnote{We refer to an instance (data sample) in the Fora corpus as a \emph{quote}. This is an excerpt from a conversation that could span one or more sentences by one speaker. See \citeauthor{schroeder_fora_2024} for more details.} from the NYC corpus as the main data set for this study, and 200 instances from the Real Talk for Change corpus as a partial replication data set. Conversation excerpts had an average length of 592 characters.

\section{Methods}

The experiment compares the annotation of 200 quotes by 5 unique annotators each according to the codebook for both NYC and RTFC corpora. First, we create a crowdworker "baseline" for the annotation task without LLM assistance for both corpora. Following typical practices, we construct a ground truth set of labels using a 3/5 majority vote for the inclusion of each label based on a set of annotations given by 5 unique annotators who independently reviewed the quote. We then test how this crowd baseline contrasts to annotations created when presenting LLM-suggested labels for annotators to review in three distinct formats in an interface. As such, in addition to our control, we adopt the following three experimental conditions testing how \textit{interface} mediates LLM-assisted annotation suggestions:

\begin{itemize}
  \item \textbf{Control: Baseline.} Annotators were not shown any LLM-generated suggestions in the annotation task.
  \item \textbf{Condition 1: Text-based suggestions.} Annotators were presented with the annotation interface, which presented the text to be annotated at the top of the screen. The text "Suggested tags:" was appended after the quote, followed by a list of LLM-generated labels, generated either by GPT-4 or Llama, according to the corpus' codebook of labels.
  \item \textbf{Condition 2: Text-based suggestions, with AI disclosure.} Same as Condition 1, except immediately following the quote, the text "Suggested tags from AI:" was appended, followed by the list of LLM-generated labels, pictured in Figure~\ref{fig:condition2}.
  \item \textbf{Condition 3: Pre-highlighted labels in interface.} The same GPT-generated label suggestions were pre-highlighted on each question in the interface, as pictured in Figure~\ref{fig:prehighlight_screenshot}.
\end{itemize}

Conditions 1-3 provide a sliding scale of assistance to an annotator. The no-assistance baseline in the Control condition provides the basis for crowd truth labels.
Text-based suggestions (Condition 1) provide some assistance, but still require the annotators to read the suggestion, then integrate the suggestions themselves into decisions for each annotation question. Condition 2 is the same as Condition 1 with the exception of including a disclosure that the suggestions were from "AI". This condition tests whether annotators would change their perception of or behavior towards suggestions if they knew their origin--- either in their perceived quality or in their rate of uptake. Finally, Condition 3 provides annotators with the strongest suggestion, drawing the annotator's attention directly to a colored highlight of the suggested label. 
These conditions were all shown to annotators through a deployment of the open source annotation interface, \href{https://github.com/davidjurgens/potato}{Potato} \cite{pei_potato_2023}. Screenshots and interface examples are in the appendix.

We ran this control and three experimental conditions on the NYC data with GPT-generated labels. In addition to this main set of experiments, we replicated Condition 1, text-based label suggestions, with labels created by by Llama 3.1 70B \cite{touvron_llama_2023}, an open-source model, using the same prompts. 
Second, we also create a no-assistance baseline control condition for the RTFC data. We use this to compare against a replication of Condition 1 (text-based suggestions created by GPT-4) on the RTFC data, testing the generalizability of our findings from a main experimental condition from the NYC data to a different codebook and dataset.

We generated label suggestions by prompting one of two LLMs. For most of our experiments, we used OpenAI's API to prompt GPT-4 \cite{openai_gpt-4_2024}, model version \verb|gpt-4-1106-preview|, zero shot, and for our replication study of Condiiton 1 on the NYC corpus, we used Llama 3.1 70b \cite{touvron_llama_2023}, accessed through the service \href{https://www.llama-api.com/}{LlamaAPI}. We used the same prompt style and instructions to prompt both GPT-4 and Llama. We prompted each model once per quote to produce a list of labels for each of the 200 quotes from the NYC and RTFC corpora. The prompt details for each task are available in the appendix.

\subsection{Survey experiment}
In order to test annotation performance at scale under a wide variety of conditions, we hired crowdworkers to complete our annotation study.
We recruited qualified annotators from \href{https://www.prolific.com/}{Prolific}. Additional details about recruitment are available in the appendix. 

Once annotators accepted the task and had read instructions, each annotator was given 20 unique annotations, with 2 randomly assigned "understanding" check questions mixed in. 200 quotes were thus annotated by 5 unique annotators in each of the controls, three experimental conditions, and our replications. Each annotator participated in just one experimental condition. Each annotator was recommended to spend 20-30 minutes on the annotation task, and spent an average of 35 minutes on the task. 


Prior to doing the task, we conducted an exercise to measure inter-annotator agreement among Prolific workers in our worker pool for this task. In order to do so, we presented 15 unique annotators with 19 unique quotes from the NYC corpus, and 20 unique annotators with 20 quotes from the RTFC corpus. We then measured inter-rater reliability on the codes for each corpus using the traditional measure of Krippendorff's alpha ($\alpha$) across annotators, which yielded low to medium levels of agreement across annotators overall, depending on the label. Overall, low to medium agreement in this context is unsurprising for both the NYC and RTFC codebooks, given the subjective, complex nature of each task. In order to form a test of task understanding, we ranked quote and label pairs by level of agreement in the IRR task. We selected these high-agreement quote and label pairs as a pool of minimum-threshold understanding questions for annotators, which we used as proxies for basic understanding of the task.
The 4 selected test questions in the NYC corpus had 13 or 14 of 15 annotators in agreement with each label, and 15-18 annotators in agreement with each label for questions pertaining to the RTFC corpus.
We included two randomly selected understanding test questions from the relevant corpus in the question bank for each annotator, which were presented in a random order within the task, and called "understanding" questions for the rest of this study.

Within the presented task, each annotator would see a quote from the dataset on the screen. Annotators were prompted to select any labels that applied to the quote, or select none if none applied. For the NYC corpus, there were 7 annotation questions for each quote, one corresponding to each top-level label. In the RTFC corpus, there were 9 annotation questions for each quote, corresponding to each top-level label. 
In the conditions where LLM-generated suggestions were presented to the annotator, we also included an additional question associated with each quote, asking annotators if suggestions were "overall", "somewhat", or "not helpful/accurate".

After being presented with all quotes, annotators were given a post-survey in order to understand their perception of the task. Before answering these self-report questions, we told them "Your answer to this question will not reflect on your performance, so please answer honestly." Annotators were given the opportunity to answer the following, rating each on a 1-5 scale, with 5 being the highest:

\footnotesize
\begin{itemize}
  \item How well do you feel you understood this labeling task?
  \item Overall, how confident do you feel in your answers on this task?
  \item After doing this task, how well do you understand the concerns and priorities of this community?
  \item After doing this task, how well could you explain this community’s concerns and needs?
\end{itemize}
\normalsize
We finished the post-survey with some demographic questions to better understand the annotator pool after their answers had already been given, including asking annotators for their race, gender, political orientation. 

\section{Results}

We compare outcomes on these annotation tasks across conditions, including the different assistance conditions, 2 corpora (NYC and RTFC), and 2 models that provided suggestions (GPT-4 and Llama).

\subsection{LLM assistance did not decrease annotation time}


\begin{table}[]
\centering
\resizebox{\columnwidth}{!}{
\begin{tabular}{l|lll}
\hline
\multicolumn{1}{|l|}{\textbf{Dataset}} & \multicolumn{1}{l|}{\textbf{Model}} & \multicolumn{1}{l|}{\textbf{Suggestion type}} & \multicolumn{1}{l|}{\textbf{Avrg. seconds}} \\ \hline
\multirow{5}{*}{NYC}                   & None                                & None (Baseline)                               & 54.3                              \\ \cline{2-4} 
                                       & \multirow{3}{*}{GPT}                & Text-based                                    &  78                              \\
                                       &                                     & Text-based, AI  disc.               & 78.8                               \\
                                       &                                     & Pre-filled                                    & 66.2                              \\ \cline{2-4} 
                                       & Llama                               & Text-based                                    & 91.1                                 \\ \hline
\multirow{2}{*}{RTFC}                  & None                                & None (Baseline)                               & 72.2                                \\
                                       & GPT                                 & Text-based                                    & 84.4                             \\ \hline
\end{tabular}
}
\caption{Average seconds spent per annotation across conditions. \tablefootnote{Most annotators were presented 20 quotations, but given some task reassignments, some completed fewer. Here, we calculate average number of seconds spent per substantive labeling question for annotators who were presented and completed at least 16 annotations in their task.}}
\label{tab:time}

\end{table}

Contrary to our pre-registered hypothesis, we found annotators in the LLM assistance conditions did not go faster than annotators in the baseline condition. To calculate time spent on the task, we calculated the average number of seconds spent on labeling an instance. There were some increases in time spent in the assistance conditions. In the assistance conditions, we included an additional short question for each annotation asking annotators to rate suggestion quality, so increases in the assistance conditions may be attributable to this additional question we asked.
Figure~\ref{fig:time} in the appendix shows time variation.
This replicates findings in \cite{overney_sensemate_2024}, which found that when qualitative coders had access to AI-generated suggestions, they actually spent longer on the annotation task than in the baseline condition.

\subsection{Assistance improves self-reported understanding of task and content}

Despite no increase in time-based productivity outcomes, annotators' self-reported experience of the task improved in most of the assistance conditions over the no-assistance controls.
Annotators self-reported higher levels of task understanding, task confidence, community understanding, and ability to explain community needs over the baseline no-assistance condition. In the NYC experiments, all assistance conditions had statistically significantly higher levels ($p < .01$, Bonferroni corrected $\alpha$ = 0.013) over the baseline condition according to a two-tailed t-test, with small to medium positive effect sizes calculated using Cohen's $D$. In the RTFC replication comparing a no-assistance control to the text-based suggestion from GPT condition, we found similar trends. The assistance condition yielded statistically significantly higher levels of task confidence, understanding of the community, and ability to explain community concerns, over the control according to a two-tailed t-test ($p < .01$) but not confidence in the task itself.


Because this is a subjective task, self-reports of understanding could theoretically increase with LLM assistance while some measure of "true" understanding on the task could decrease. To test this, we analyzed attention in the NYC control and assistance conditions. The failure rate of understanding checks across all conditions was 11.6\%, and we did not find a statistically significantly change across conditions when annotators had access to AI assistance. This may provide at least basic assurance that providing assistance did not immediately elicit overreliance on the assistance to the point the basic task was not understood. 


\subsection{Annotators overwhelmingly like and take the suggestions}

To calculate annotators' rated helpfulness of the suggestions, we converted their ratings of suggestion helpfulness and accuracy into numeric values as follows: "Overall helpful/accurate" to a 2, "Somewhat helpful/accurate" to a 1, and "Not helpful/accurate" to a 0. Across conditions, LLM suggestions were rated as between somewhat and very helpful (mean: \( 1.24\)). We did not find notable differences in ratings of helpfulness between GPT-4 and Llama, or in the way suggestions were presented (Condition 1: text-based, Condition 2: text-based + AI disclosed, and Condition 3: pre-filled). Helpfulness of label suggestions was also rated similarly in the NYC and RTFC annotation tasks, suggesting no one model worked better than the other, and the assistance was helpful in two different labeling contexts.

Reflecting this perceived helpfulness and accuracy, we observed stronger overlap between the set of human crowd annotations and LLM annotations when the humans had been exposed to LLM suggestions. To observe this, we created a human crowd label for each quote. Crowd decisions for labels were made by checking if, for each label, the label was assigned to the quote by at least 3 (of the 5\footnote{In some cases, more than 5 unique annotators provided labels for the quote, in which case 5 unique annotators' judgements were randomly sampled to use as crowd judgements. In a few cases per condition, fewer than 5 unique annotators' judgments could be obtained for each quote. Of 200 quotes, three quotes from the baseline condition, 7 from Condition 2, 2 from Condition 3, and 16 from the Llama replication of Condition 1 had four annotators, and were dropped from subsequent crowd analyses predicated on 5 unique annotators.}) annotators, giving a final set of labels where, for each, at least 3 annotators agreed on the relevance of the label. We repeated this process at a crowd decision threshold of 4 annotators, as well as full consensus of all 5 annotators. 
To contrast a human crowd baseline to the set of LLM annotations for a given quote, we obtained the set intersection of labels applied by the LLM to a particular quote  and the set of labels applied by a crowd decision of human annotators to that same quote. This metric captures the proportion of LLM-suggested tags also given by the human crowd under a given condition \( c \) and approval threshold \( \theta \), and is further detailed in the appendix.

\begin{figure}[ht]
\centering
\includegraphics[width = 200pt]{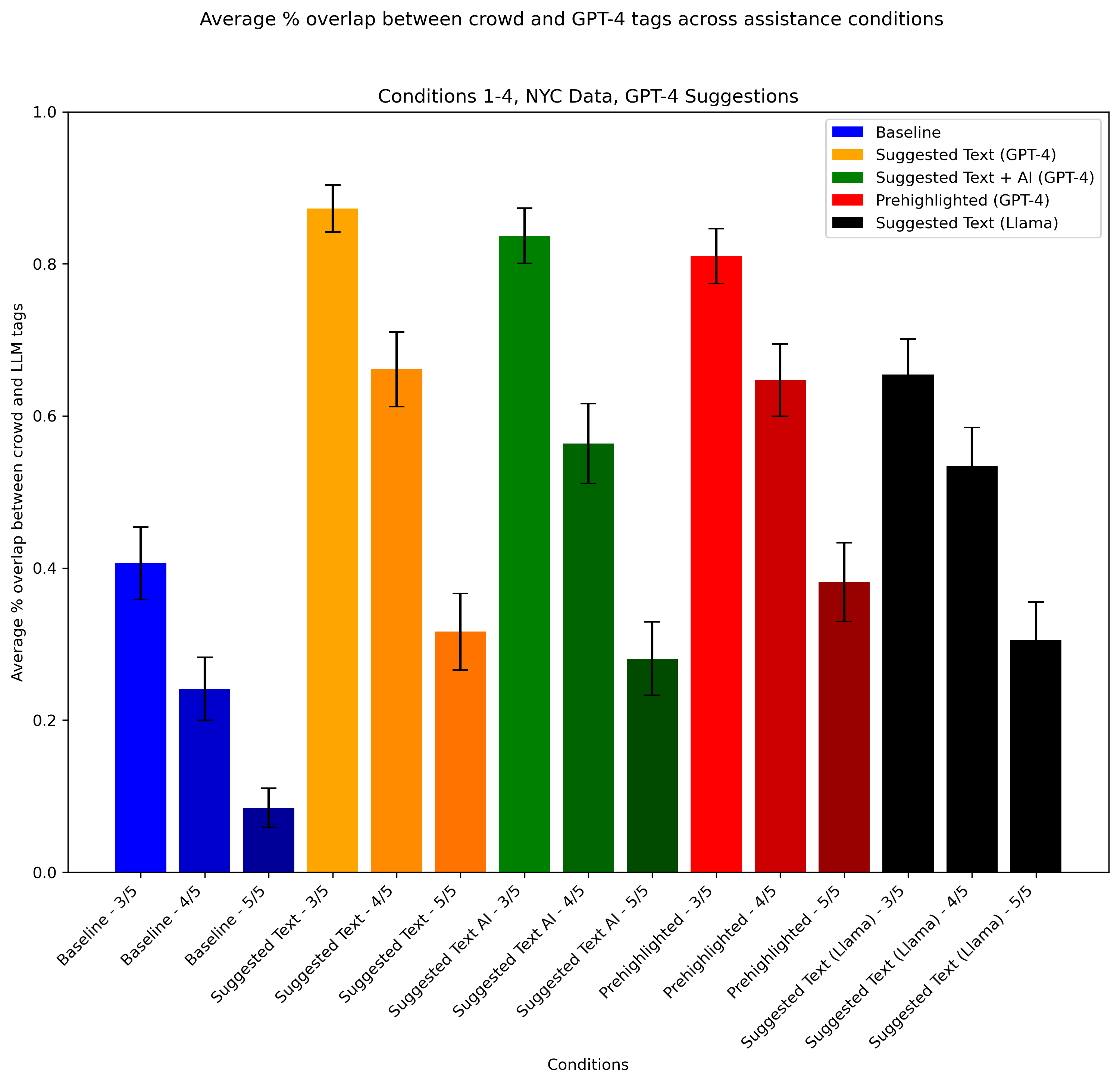}
\caption{Average ratio of overlap between crowd annotator labels with LLM-suggested labels by condition, and crowd decision threshold, and source of LLM suggestions.}
\label{fig:uptake}
\end{figure}

Treating crowd decisions on the unassisted baseline condition for the NYC corpus as ground truth, just 40\% of labels given by a crowd decision of 3 annotators overlapped with the GPT-4 suggestions. This further dropped sharply to  24\% when the crowd threshold is raised to 4 annotators, and to just 8\% when raised to full consensus of 5 annotators. Figure~\ref{fig:uptake} shows that in all LLM assistance conditions, run for Conditions 1, 2, and 3 with GPT-4 labels on NYC data, the overlap between the crowd ground truth created with LLM assistance and the LLM label set increased dramatically.
We display this in terms of different crowd decision thresholds. At a crowd decision threshold of 3/5 and looking at text-based suggestions from GPT, crowd labels had an average overlap ratio between crowd labels and suggested labels of 81-87\% depending on the presentation of suggestions, 56-65\% at a crowd decision threshold of 4, and between 28-38\% for full crowd consensus of 5. In other words, overlap with LLM suggestions increased as much or more than 40\% at the typical decision threshold of 3 when human annotators were given these suggestions to review, a statistically significant increase according to a two-tailed t test at $p = .05$. Text-based Llama suggestions resulted in similar results for the NYC corpus, but had lower overlap (.65 at crowd threshold of 3, .53 at 4, and .3 at 5).  

We also observed consensus agreement of all 5 annotators was significantly more likely in Condition 3, where suggestions from GPT were presented most strongly by appearing pre-filled in the interface as highlighted labels.
We observed that full consensus, or full agreement by all 5 annotators, increased from just 8\% in the no assistance baseline to 38\%  in the pre-highlighted label condition. Using a two-tailed t-test, we find this is a statistically significant increase a $p = .001$, including Bonferroni correction for multiple comparisons. This suggests the interface used to present suggestions to annotators can impact the strength of suggestion uptake.

\subsection{Using human-reviewed, LLM-assisted labels as ground truth significantly inflates reported model performance}

Using LLMs to annotate or augment annotations that can be used to train models or evaluate model performance is tempting, given the challenge of scaling annotation, particularly for subjective tasks that may be challenging for crowd workers.
How much does model performance appear to improve on these subjective tasks when we use LLM-assisted annotations, even when reviewed by many humans, as ground truth? 

To examine this, we first created baseline ground truth labels for the 20 NYC labels by aggregating annotations made for the quote by 5 unique annotators into a 3/5 majority vote. For each of the 20 labels in the codebook, we calculate an F1 score of "model performance" using the human crowd labels as ground truth, testing either the labels for that quote suggested by GPT or Llama labels for the entire set of 200 NYC quotes, calculating an F1 score of model performance for that label.
Treating our annotations from our control condition (aggregated into 3/5 majority votes) as ground truth shows overall low LLM performance on these labeling tasks, with similar performance for both GPT-4 and Llama: the average weighted F1 score of GPT-4 performance across all labels is .47 ($\sigma = .18$) when using the human crowd labels as ground truth, and .44 ($\sigma = .18$) for Llama when using the human crowd labels as ground truth. Interestingly, when GPT-4 labels for the NYC corpus are used as ground truth labels and compared to Llama labels, the average F1 score is significantly higher at .62 ($\sigma = .10$) compared to performance when using a human crowd baseline. The individual breakdown of label-level performance is available in Table~\ref{tab:performance}, and breakdowns of GPT-4 versus Llama are available in the appendix in Table~\ref{tab:gpt_llama_head_to_head}.

We next sought to compare the calculated F1 score on a per-label basis when using the crowd labels from the control condition to labels generated by annotators who reviewed LLM suggestions. We aggregated labels from annotators reviewing text-based suggestions from GPT-4 ("GPT-assisted") or Llama ("Llama-assisted"). We used the same 3/5 majority to approve each label, constructing a new ground truth condition for annotations made with assistance. When using the GPT-assisted ground truth, the average weighted F1 score of GPT-4 performance across all labels increased to .79 ($\sigma = .14$), for an average increase in F1 score of +.32. When using crowd-aggregated "Llama-assisted" labels as ground truth, the average weighted F1 score of Llama performance across all labels increased to .79 ($\sigma = .16$), for a similar average increase in F1 score of +35. Performance on some labels increased by substantially more than that average, including "Role of community organizations: Trust, Rapport, \& Relationships", which increased from .28 to .82  when GPT-4's labeling performance was evaluated on GPT-assisted ground truth labels. 

\begin{table*}[]
\centering
\tiny
\renewcommand{\arraystretch}{1.3}  
\begin{tabular}{llp{1.5cm}p{1.2cm}p{1.2cm}p{1.4cm}p{1.4cm}}
\cline{4-7}
                                                                             &                                        & \multicolumn{1}{l|}{}                        & \multicolumn{4}{c|}{\textbf{\rule{0pt}{2.5ex}Ground truth v. LLM}, weighted F1 score}                                                     \\ \hline
\multicolumn{1}{|l|}{{\color[HTML]{000000} \textbf{\rule{0pt}{2.5ex}Top-level label}}}        & \multicolumn{1}{l|}{\textbf{Sublabel}}               & \multicolumn{1}{l|}{\textbf{\raggedright Frequency}} & \textbf{\raggedright Human crowd \\ v. GPT labels} & \textbf{Human crowd v. Llama labels} & \textbf{GPT-assisted human labels vs GPT labels} & \textbf{Llama-assisted human labels vs Llama labels}\\ \hline
& Civic Organizations    & 30 & 0.57     & 0.47    & 0.87 {\color[HTML]{008000} +.30}   & 0.85  {\color[HTML]{008000} +.38}    \\
& Employers     & 8  & 0.50    & 0.50         & 0.91 {\color[HTML]{008000} +.41}  & 0.96  {\color[HTML]{008000} +.46}                                                   \\
& Family \& Friends           & 29         & 0.58           & 0.60      & 0.78 {\color[HTML]{008000} +.20}      & 0.83 {\color[HTML]{008000} +.23}          \\
& Health Care Providers   & 16      & 0.50        & 0.54    & 0.86 {\color[HTML]{008000} +.36}       & 0.77  {\color[HTML]{008000} +.23}                                                   \\
\multirow{-5}{*}{\textbf{\rule{0pt}{2.5ex}External motivations}}    & Social \& News Media  & 5      & 0.38     & 0.42    & 0.77 {\color[HTML]{008000} +.39}    & 0.96 {\color[HTML]{008000} +.54} \\ \hline
& Discussion of post-pandemic future         & 15         & 0.81       & 0.62        & 0.81 {\color[HTML]{808080} +.0}    & 0.84 {\color[HTML]{008000} +.22}           \\
\multirow{-2}{*}{\textbf{\rule{0pt}{2.5ex}Future visions \& Takeaways}}     & Reflections on the conversation    & 9      & 0.13    & 0.0           & 0.40 {\color[HTML]{008000} +.27}   & 0.28 {\color[HTML]{008000} +.28} \\ \hline
& Basing decisions on data   & 4      & 0.36     & 0.33        & 0.81 {\color[HTML]{008000} +.45}  & 0.86 {\color[HTML]{008000} +.53}                                                    \\
& Getting back to normal    & 16        & 0.47        & 0.41         & 0.85  {\color[HTML]{008000} +.38}       & 0.73   {\color[HTML]{008000} +.32}     \\
\multirow{-3}{*}{\textbf{\rule{0pt}{2.5ex}Intrinsic motivations}}  & Not wanting to get the virus  & 21 & 0.44    & 0.46       & 0.89 {\color[HTML]{008000} +.45}         & 0.96 {\color[HTML]{008000} +.50}                                                   \\ \hline
{\color[HTML]{000000} }       & Resilience, Connection, \& Hope   & 23     & 0.42   & 0.23       & 0.62  {\color[HTML]{008000} +.20}    & 0.62 {\color[HTML]{008000} +.39}   \\
\multirow{-2}{*}{{\color[HTML]{000000} \textbf{\rule{0pt}{2.5ex}Personal COVID experience}}} & {\color[HTML]{000000} Stress, Fear,  \& Uncertainty} & 50     & 0.67   & 0.57    & 0.88 {\color[HTML]{008000} +.21}   & 0.84  {\color[HTML]{008000} +.27}        \\ \hline
{\color[HTML]{000000} }   & Significant Impact Resources that helped             & 30   & 0.63     & 0.67  & 0.85  {\color[HTML]{008000} +.22}  & 0.82{\color[HTML]{008000} +.15}                                                    \\
\multirow{-2}{*}{{\color[HTML]{000000} \textbf{\rule{0pt}{2.5ex}Resources that helped}}}  & Unmet community needs  & 22     & 0.45    & 0.42    & 0.55 {\color[HTML]{008000} +.10}   & 0.76 {\color[HTML]{008000} +.34}   \\ \hline
& Health Education \& Support  & 13   & 0.20     & 0.32     & 0.76  {\color[HTML]{008000} +.56}  & 0.61 {\color[HTML]{008000} +.29}  \\
& Incentives     & 4   & 0.30   & 0.32   & 0.63 {\color[HTML]{008000} +.33}  & 0.67 {\color[HTML]{008000} +.35}  \\
& Reducing barriers   & 6   & 0.32  & 0.36 & 0.78 {\color[HTML]{008000} +.46} & 0.90  {\color[HTML]{008000} +.54}       \\
\multirow{-4}{*}{\textbf{\rule{0pt}{2.5ex}Role of community organizations}}  & Trust, Rapport, \& Relationships        & 10   & 0.28   & 0.29    & 0.82 {\color[HTML]{008000} +.54}       & 0.67 {\color[HTML]{008000} +.38}     \\ \hline
& Did not vaccinate          & 4 & 0.73  & 0.53    & 0.93 {\color[HTML]{008000} +.20}   & 0.90  {\color[HTML]{008000} +.37}                                                  \\
\multirow{-2}{*}{\textbf{\rule{0pt}{2.5ex}Vaccine hesitancy}}
& Mistrust or Skepticism & 19     & 0.65   & 0.80     & 0.94 {\color[HTML]{008000} +.29}  & 0.94 {\color[HTML]{008000} +.14}                                                   \\ \hline
\end{tabular}  
\caption{Performance evaluation of human crowd labels against GPT labels and Llama labels, then GPT-assisted human labels against GPT labels and Llama-assisted human labels against Llama labels. \textbf{Frequency} refers to $n$ observations found across crowd-aggregated annotations on the 200 quotations from the NYC corpus at the crowd threshold of 3/5. Performance increases an average of + .32 when using GPT-assisted crowd labels against GPT labels compared to an unassisted crowd baseline, and + .35 when using Llama-assisted crowd labels against Llama labels compared to an unassisted crowd baseline.}
\label{tab:performance}
\end{table*}



\section{Discussion}

While mainstream perceptions suggest LLMs can help speed up annotation, we find that if humans review individual LLM-generated suggestions, annotation time does not decrease. 
From our own measures, we find annotators who were given assistance self-reported improved task understanding. Baseline  task understanding, as measured through test questions, remained constant across conditions. Follow-up work could examine whether seeing AI suggestions sometimes increased task time because there was more information to process, and if suggestions may teach new annotators to do a complex task, increasing their confidence when assistance is given. Future work can also examine whether annotators under a different compensation incentive structure approve LLM suggestions more quickly than these workers did.


Annotation is usually the first step in creating ground truth data for evaluating a model's performance on an NLP task. Complex, subjective tasks are common in the subfield of NLP for computational social science and cultural analytics. Using LLMs to annotate data used for training and evaluation of a task is attractive due to time and cost efficiency compared to hiring humans, who may noisily interpret a subjective task like this one. However, our findings provide a cautionary note: humans often take LLM suggestions they are given, even when a human individually reviews each label that contributes to the ground truth. As such, using LLM annotations, or LLM-assisted labels like those described here, in the creation of ground truth could inflate measures of LLM performance on these subjective annotation tasks. 

In taking LLM suggestions, annotators homogenize "ground truth" on these tasks towards LLM baselines. In NLP for CSS tasks and in qualitative coding, using labels to measure prevalence of a concept in text is common, so LLM suggestions can change these measurements. For example, imagine using these annotations to analyze cited intrinsic motivations for vaccine decision-making given by participants in the NYC conversations: we could view the list of most commonly cited \textit{Intrinsic motivation} labels to understand the most common motivations participants have for vaccine decisions. In the human crowd baseline, \textit{Intrinsic motivations: Basing decisions on data} was the least commonly occurring label, whereas when crowd labels were generated by annotators under Condition 1 (GPT-influenced labels presented in a text format), it was nearly five times more common, tying \textit{"Getting back to normal"} as the second most common \textit{Intrinsic motivation} tag. Analysts using LLM-assisted annotations for a substantive analysis of the conversations could thus come to a different conclusion about the relative importance of \textit{Basing decisions on data} as an intrinsic motivation for vaccine uptake in this community: in the case of the human crowd baseline, it was rare (4 instances), and in the LLM-influenced Condition 1, it became a heavily assigned \textit{Intrinsic motivations} tag, with 19 occurrences in the data. 

Homogenization towards an LLM baseline may not be an inherent problem, but the low performance we observe from both GPT-4 and Llama when contrasted with a human baseline raises several possible explanations that deserve attention when planning LLM annotation tasks. Depending on the level of subjectivity, or even potential polarization of the construct being annotated, human annotators may be noisier in interpreting the annotation task and providing their judgments. However, in situations where humans have high agreement, but LLMs systematically annotate the same construct differently, the LLM may be using a different background concept \cite{jacobs_measurement_2021} for annotating the labels than human crowdworkers do. For example, given a starting F1 score of just .2 comparing LLM annotations to the human crowd baseline, GPT-4 may operationalize the identification \textit{Health Education and Support} differently than our aggregation over human crowdworkers did. When LLM annotations are used to identify this construct, annotations may thus be measuring something different than the humans crowd baseline does, and the concept of \textit{Health Education and Support} being annotated becomes \textit{more} like the LLM's conception of \textit{Health Education and Support} when annotators have its assistance. 

LLM assistance also inherently increases measures of interrater reliability when the same LLM suggested are given to multiple annotators. IRR measures are used as a positive signal of reliability when humans annotate social science concepts \cite{mcdonald_reliability_2019}. However, when annotators use LLM assistance and IRR is still used as a positive signal of reliability, researchers would need to be confident that the LLM's suggestions--- and the way it operationalizes each background concept being identified--- is \textit{more correct} than human judgments, since humans are likely to adopt LLM suggestions. In subjective tasks, this can be hard to verify and prove, but error analysis of specific ambiguous labels may help.

Second, there are many tasks in NLP where varied annotator perspective can be valuable \cite{basile_toward_2023, plank_problem_2022}, both from the perspective of constructing a robust ground truth \cite{aroyo2013crowd, yan_learning_2014}, or for representing diverse human perspectives on a complex social construct like hate speech \cite{sap_annotators_2022}. In qualitative research, some traditions explicitly embrace divergent annotator perspective in order to widen insight in qualitative annotation,  \cite{mcdonald_reliability_2019} so in these cases, homogenization of insight as a result of LLM involvement in annotation may be seen as a source of concern \cite{schroeder_large_2025}. In both the cases of creating ground truth for NLP tasks and qualitative annotation, practitioners should know that using or providing LLM assistance to annotators will likely result in lessened variation.

Given potential consequences for representation and construct validity that vary by task, researchers should only proceed with LLM-assisted annotation with a level of caution appropriate to their task and goal. They should recognize that using LLM assistance to construct ground truth labels inflates perceptions of model performance on that task, even when humans review them and individual judgments are aggregated into a crowd ground truth.
Follow-up work can investigate if these findings hold for a similarly complex annotation task, a less complex annotation task, and when employing expert annotators rather than unspecialized crowd workers. Homogenization effects may be lessened by presenting information about model confidence, or alternative ideas, and not just a single set of suggestions. 

\section{Conclusion}

In a pre-registered experiment with 410 unique annotators and 7,000+ annotations across 3 experimental conditions, 2 models, and 2 datasets, we find that presenting crowdworkers with LLM-generated annotation suggestions did not make them faster, but did improve their self-reported confidence in the task. Annotators strongly uptook suggestions, changing the label distribution to more closely resemble the LLM's proposed distribution.

We also found that using LLM-assisted labels to evaluate model performance resulted in much higher reported F1 scores than when using a human crowd baseline, with increases in F1 scores for model performance on some labels by as much as +x.56. Obviously, using labels influenced by the model to evaluate the model is not standard or advisable in classic evaluation paradigms. However, in the many systems being created that "just put a human in the loop" to review LLM annotation outputs, this paradigm of reviewing LLM outputs to "approve" them is increasingly likely to occur. Practitioners should know that, especially in subjective tasks, simply reviewing LLM suggestions will nudge the distribution of label outputs towards an LLM baseline, even if humans are given a chance to review the outputs. 

\section{Limitations}

Crowdworkers may be particularly susceptible to this kind of influence from LLM suggestions, however, their continued employment as a standard for annotation in the field justifies their employment in the task here on a deliberately ambiguous task. 
Furthermore, different results may be reached with specialized annotators with particular domain knowledge. Annotators with a relationship to the data, including a relationship to the community of interest, may also shape how they align with or reject AI interpretations of the data. Follow-up work can investigate whether domain experts have less anchoring bias than we observed here, and whether they confidently defect from LLM suggestions when needed.

\section{Acknowledgements}
We appreciate the feedback from members of the Center for Constructive Communication and Cortico as this work was developed, and in particular, we thank Dimitra Dimitrakopoulou for her feedback and support. We thank Jiaxin Pei for his consultation on Potato, and development work on Potato. We thank David Rand and his lab for their feedback on this work. 
We thank Erin Kim for research assistance on the literature review for this piece. 

\bibliography{custom} 

\appendix

\section{Calculating overlap metric}

In Section 5.3, we briefly describe the way we calculate overlap between a set of crowd labels and LLM labels. We further describe it here:

Let:
\begin{itemize}
  \item \( i \in \{1, 2, \ldots, N\} \): be the ID of the data sample, here, a quote from one of the two corpora
  \item \( c  \) be the experimental condition, such as "control" or "text-based Llama suggestion" (Condition 1)
  \item \( \theta \in \{3, 4, 5\} \): crowd agreement threshold (number of unique annotators who must agree on a tag)
  \item \( A_i \): set of AI-suggested tags for instance, \( i \) generated either by GPT or Llama, depending on the condition
  \item \( H_{i}^{(c,\theta)} \): set of human-submitted tags for instance \( i \), under condition \( c \), using crowd threshold \( \theta \)
\end{itemize}

Using these definitions, the \textbf{intersection ratio} for except \( i \), condition \( c \), and threshold \( \theta \) is:

\[
R_i^{(c, \theta)} = 
\begin{cases}
\frac{|A_i \cap H_{i}^{(c, \theta)}|}{|A_i|}, & \text{if } |A_i| > 0 \\

\end{cases}
\]

The \textbf{average intersection ratio} across all \( N \) instances for a given \( c \) and \( \theta \) is:


\begin{align*}
\bar{R}^{(c, \theta)} &= \frac{1}{N'} \sum_{i=1}^{N} R_i^{(c, \theta)} \\
\text{where } N' &\text{ is the number of instances with defined } \\ R_i^{(c, \theta)}
\end{align*}

This metric captures the proportion of LLM-suggested tags also given by the human crowd under a given condition \( c \) and approval threshold \( \theta \).

\section{Corpus details}

Corpora were selected based on the availability of both public conversation data and the existence of a human-created codebook for annotation which was shared with us by our collaborating organizational partner. The Fora corpus is lightly anonymized, with speaker names removed. For the NYC and RTFC sample of 200 quotations, we manually reviewed and removed any personally identifiable information before using it in a prompt to either model, and before showing it to annotators. Participants in the NYC and RTFC conversation collections were aware their voices would be collected and used to inform the public about issues in their community, as well as potentially used in research. Both the NYC and RTFC conversations contain a mix of standard American English and African American English, as well as Spanish, though less often. We eliminated any participant comments in Spanish prior to sampling 200 excerpts for use this study, given that performance on this tasks could not be compared across languages, and the codebook was developed in English. Given the unreliability of African American English dialect detectors as of the time of this submission, particularly on transcribed speech, we are not able to estimate prevalence of African American English or other dialects that may be in this corpus. Demographic information of speakers was not collected in the NYC or RTFC conversation corpora, but more details on the corpora are available in the Fora corpus, and we can provide annotator demographics with access requests.

\section{Additional results}

\begin{figure*}[ht]
\centering
\includegraphics[width = 400pt]{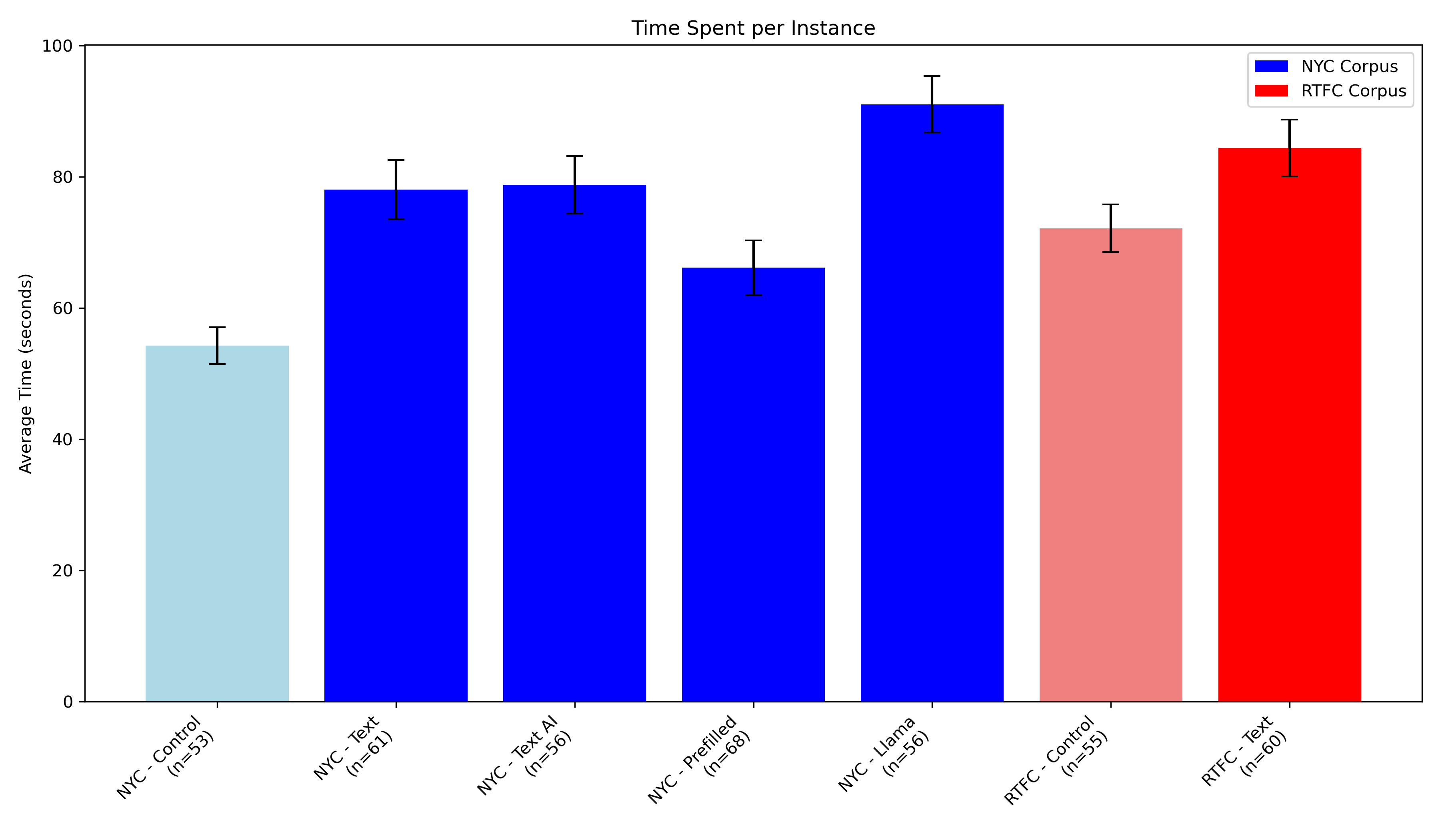}
\caption{Time (number of seconds) spent by annotators on the task, for Control and Conditions 1-3 of NYC experiments, including the replication of Condition 1 with Llama. In red on the right, Control and Condition 1 replications on the RTFC data. $N$ in each condition indicates the number of annotators who completed a task with at least 16 assigned quotes for labeling.}
\label{fig:time}
\end{figure*}

In addition to Table \ref{tab:time}, we provide a visualization of time taken on control and Conditions 1-3 (including GPT and Llama replication of Condition 1) on the NYC corpus in Figure \ref{fig:time}.

\begin{figure*}[ht]
  \centering
  \includegraphics[width=\textwidth]{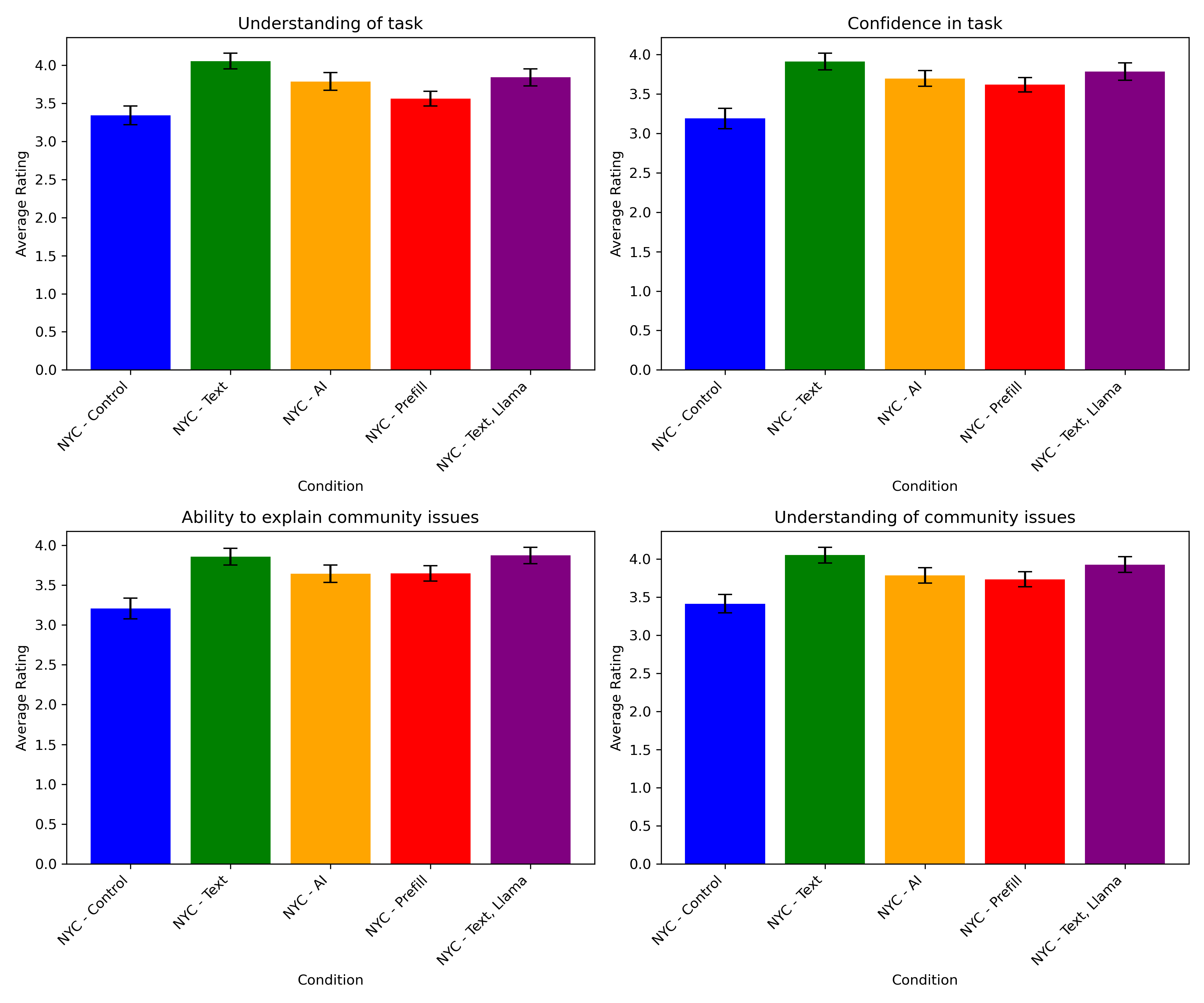}
  \caption{Four self-reported measures across conditions.}
  \label{fig:double_figure}
\end{figure*}

Figure \ref{fig:double_figure} shows annotator self-reports from post-surveys following performing the annotation task. Annotators with all forms of AI assistance self-report slightly higher levels of task understanding and confidence, as well as ability to explain community issues and an understanding of the community. On the $x$ axis, the condition is noted. The $y$ axis denotes the average self-reported value on a 1-5 likert scale.

\subsection{Additional results}

\begin{table*}
\centering
\tiny
\begin{tabular}{llllll}
\cline{4-6}
                                                                             &                                                      & \multicolumn{1}{l|}{}                        & \multicolumn{3}{c|}{\textbf{GPT ground truth, Llama test}} \\ \hline
\multicolumn{1}{|l|}{{\color[HTML]{000000} \textbf{Top-level label}}}        & \multicolumn{1}{l|}{\textbf{Sublabel}}               & \multicolumn{1}{l|}{\textbf{Frequency (n)}} & \textbf{F1}    & \textbf{Precision}    & \textbf{Recall}   \\ \hline
                                                                             & Civic Organizations                                  & 25                                           & 0.60           & 0.51                  & 0.84              \\
                                                                             & Employers                                            & 24                                           & 0.50           & 0.81                  & 0.54              \\
                                                                             & Family \& Friends                                    & 24                                           & 0.52           & 0.63                  & 0.79              \\
                                                                             & Health Care Providers                                & 21                                           & 0.53           & 0.55                  & 0.76              \\
\multirow{-5}{*}{\textbf{External motivations}}                              & Social \& News Media                                 & 14                                           & 0.32           & 0.47                  & 0.57              \\ \hline
                                                                             & Discussion of post-pandemic future                   & 25                                           & 0.73           & 0.88                  & 0.60              \\
\multirow{-2}{*}{\textbf{Future visions \& Takeaways:}}                      & Reflections on the conversation                      & 4                                            & 0.15           & 0.43                  & 0.75              \\ \hline
                                                                             & Basing decisions on data                             & 12                                           & 0.42           & 0.50                  & 0.75              \\
                                                                             & Getting back to normal                               & 13                                           & 0.38           & 0.64                  & 0.69              \\
\multirow{-3}{*}{\textbf{Intrinsic motivations:}}                            & Not wanting to get the virus                         & 58                                           & 0.48           & 0.78                  & 0.69              \\ \hline
{\color[HTML]{000000} }                                                      & Resilience, Connection, \& Hope                      & 20                                           & 0.32           & 0.47                  & 0.35              \\
\multirow{-2}{*}{{\color[HTML]{000000} \textbf{Personal COVID experience:}}} & {\color[HTML]{000000} Stress, Fear,  \& Uncertainty} & 55                                           & 0.66           & 0.65                  & 0.80              \\ \hline
{\color[HTML]{000000} }                                                      & Significant Impact Resources that helped             & 33                                           & 0.64           & 0.53                  & 0.82              \\
\multirow{-2}{*}{{\color[HTML]{000000} \textbf{Resources that helped}}}      & Unmet community needs                                & 11                                           & 0.42           & 0.56                  & 0.45              \\ \hline
                                                                             & Health Education \& Support                          & 24                                           & 0.20           & 0.54                  & 0.63              \\
                                                                             & Incentives                                           & 15                                           & 0.28           & 0.48                  & 0.73              \\
                                                                             & Reducing barriers                                    & 22                                           & 0.35           & 0.41                  & 0.59              \\
\multirow{-4}{*}{\textbf{Role of community organizations}}                   & Trust, Rapport, \& Relationships                     & 18                                           & 0.20           & 0.63                  & 0.67              \\ \hline
                                                                             & Did not vaccinate                                    & 12                                           & 0.60           & 0.86                  & 0.50              \\
\multirow{-2}{*}{\textbf{Vaccine hesitancy}}                                 & Mistrust or Skepticism                               & 28                                           & 0.57           & 0.88                  & 0.82              \\ \hline
                                                                             &                                                      &                                              &                &                       &                   \\
                                                                             &                                                      &                                              &                &                       &                  
\end{tabular}
\caption{This table compares GPT-4 and Llama annotations on the NYC corpus to each other, by treating GPT-4 labels as ground truth (giving the value in the Frequency column) and scoring Llama annotations against GPT-4. The average F1 score for Llama against the GPT-4 baseline is .62 ($\sigma = .10$), which was much higher than either model's comparison to human baseline, discussed in the main body of the paper.}
\label{tab:gpt_llama_head_to_head}
\end{table*}

\clearpage
\onecolumn
\section{Interface screenshots}

\begin{figure}[ht]
\centering
\includegraphics[width = 400pt]{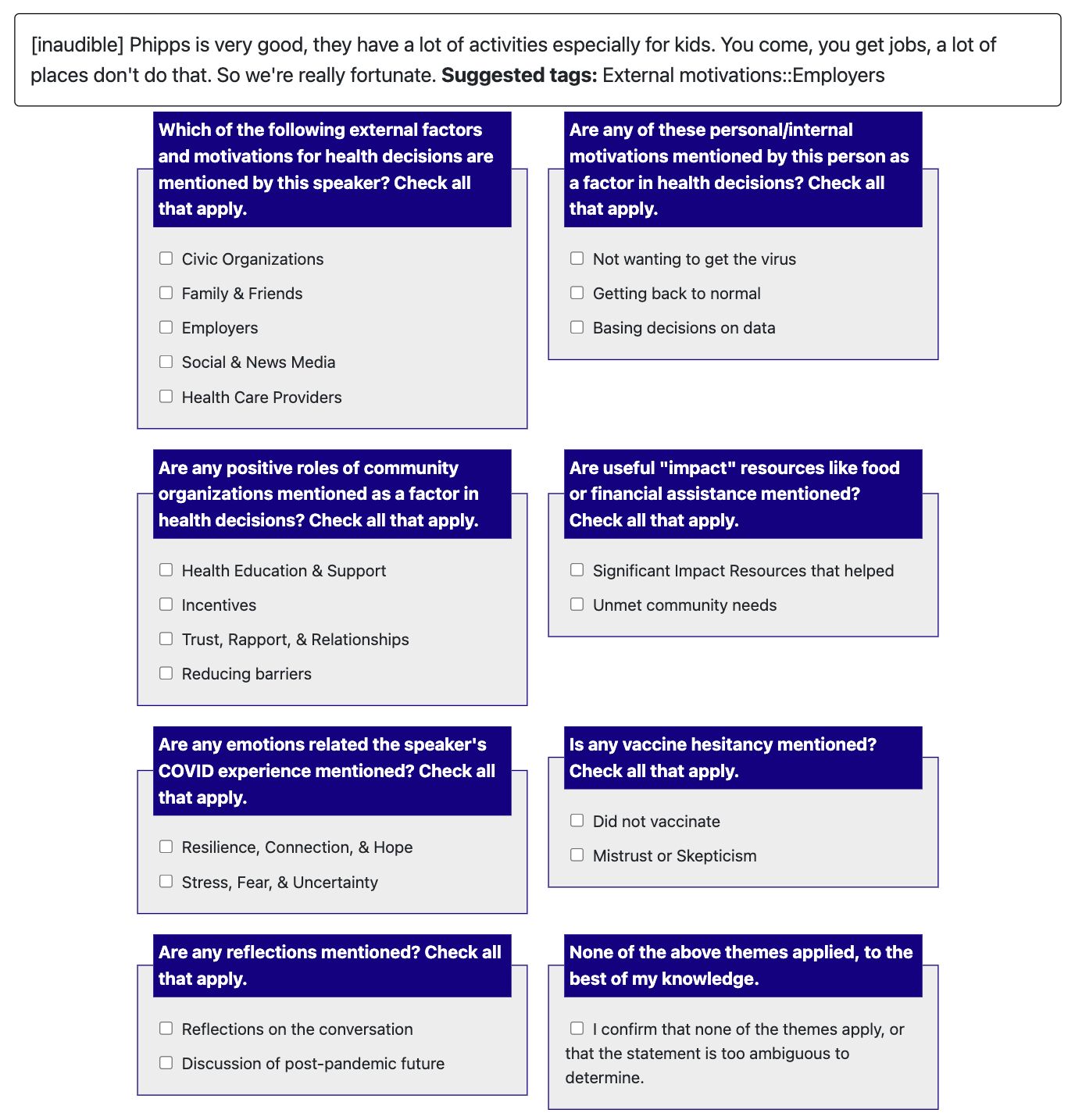}
\caption{\textit{Condition 1}: Text-based label suggestions in interface, with example and label suggestions from NYC corpus. Interactive tooltips gave extended code definitions, and the full codebook was linked in the header of the annotation task. Annotators viewed one additional question rating suggestion quality for the quote at the bottom of the page before advancing.}
\label{fig:condition2}
\end{figure}

\newpage

\begin{figure}[ht]
\centering
\includegraphics[width = 300pt]{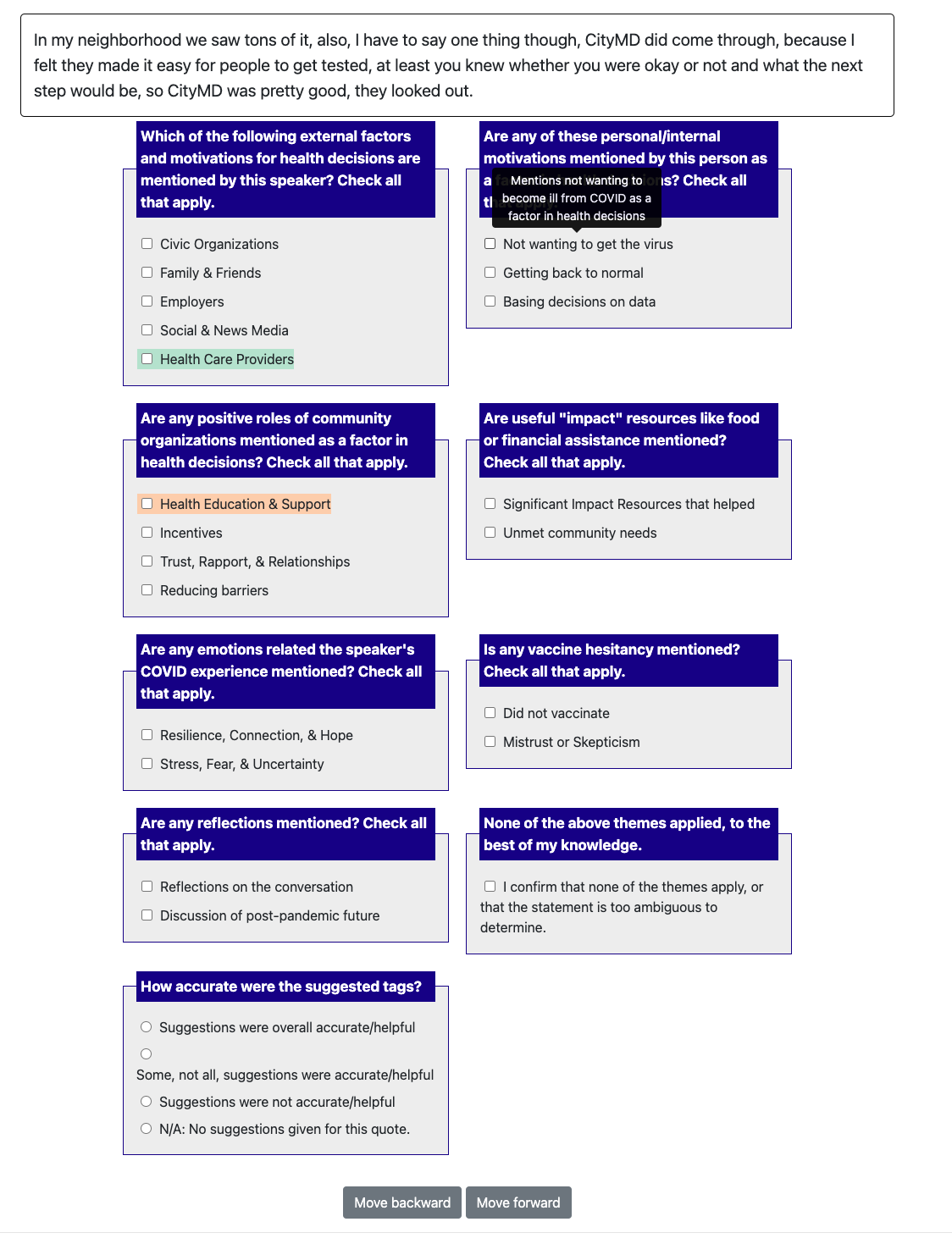}
\caption{\textit{Condition 3:} Pre-highlighted label suggestions in interface, with example and label suggestions from NYC corpus showing highlighted suggestions included "Health Care Providers" and "Health Education \& Support". Interactive tooltip with code definition for a different code is shown activated with a code definition displayed in upper right.}
\label{fig:prehighlight_screenshot}
\end{figure}

\newpage
\clearpage

\section{NYC Conversation Corpus Details}
\subsection{Conversation guide}

The NYC Public Health Corps developed a codebook in partnership with Cortico to elevate concerns from community members during the pandemic period. The main questions asked of participants in each conversation were:

\begin{itemize}
    \item Opportunities \& Challenges of Resourcing
    \begin{itemize}
        \item What COVID-19 resources have been most helpful for you during the pandemic? And, why?
        \item What challenges did you have finding and using resources intended to help you with COVID-19?
    \end{itemize}
    \item Vaccine Experiences and Decision Making
    \begin{itemize}
        \item Can you describe a key moment that influenced your decision about the COVID-19 vaccine? 
        \item Reflecting on this experience you just shared, what information or circumstances helped you make that decision? 
    \end{itemize}
    \item Role of Community Organizations in Community
    \begin{itemize}
        \item Can you share a story or experience about how the community organizations in your neighborhood supported you and your community during the pandemic?
        \item How did those community organizations impact decisions related to COVID-19 vaccinations in your neighborhood?
    \end{itemize}
    \item Future Resources
    \begin{itemize}
        \item What will a post-pandemic future look like in your community? 
        \item How can community organizations help your community thrive in the future?
    \end{itemize}
\end{itemize}
\newpage
\subsection{NYC Codebook}

\begin{enumerate}
    \item \textbf{External motivations} (Theme)
    \begin{itemize}
        \item \textbf{Civic Organizations:} The speaker mentions civic organizations like non-profits, churches, NGOs, and community clubs, associations as a factor in health decisions
        \begin{itemize}
            \item Example: "My church helped me find alternative childcare during the pandemic."
        \end{itemize}
        \item \textbf{Family \& Friends:} Mentions family and friends as a factor in health decisions
        \begin{itemize}
            \item Example: "My mom was really opinionated about this from the start. She really wanted us to get the vaccine. She even helped drive us to the vaccine clinic because I don't own a car."
        \end{itemize}
        \item \textbf{Employers} Mentions place of employment as a factor in health decisions, including employer providing resources or opportunities for vaccination, or co-workers setting a model of vaccine behavior as a factor in health decisions
        \begin{itemize}
            \item Example: "My job offered drop-in vaccine clinics, which was helpful since the regular clinic hours happen during my work hours."
        \end{itemize}
        \item \textbf{Social \& News Media:}  Mentions social media or news media as a factor in health decisions, including Facebook, Twitter, etc. Traditional network, local news, newspapers, or print media that is online
        \begin{itemize}
            \item Example: "A lot of my friends were posting on Facebook about the vaccine having a chip. That made me nervous."
        \end{itemize}
        \item \textbf{Health Care Providers:} Mentions healthcare providers as a factor in health decisions, including doctors or nurses
        \begin{itemize}
            \item Example: "My doctor gave me some information, but I don't trust that the information was up to date and I still don't want the vaccine. So no I haven't gotten it."
        \end{itemize}
    \end{itemize}
    \item  \textbf{Intrinsic Motivations} (Theme)
    \begin{itemize}
        \item \textbf{Fear of Virus:} Mentions not wanting to become ill from COVID as a factor in health decisions
        \begin{itemize}
            \item Example: "My church helped me find alternative childcare during the pandemic."
        \end{itemize}
        \item \textbf{Getting Back to Normal:} Mentions a desire for a return to activities and social routines as a factor in health decisions
        \begin{itemize}
            \item Example: "I thought you know what, if this can help us just keep the kids in school then I'll get the vaccine even if I hate it."
        \end{itemize}
        \item \textbf{Basing decisions on data:} Mentions considering data, research, or evidence when choosing to get vaccinated
        \begin{itemize}
            \item Example: "I was seeing these studies show that there is an increased chance of heart problems after the vaccine. So I did not want to get it because I have heart issues in my family."
        \end{itemize}
    \end{itemize}
    \item \textbf{Role of Community Organizations} (Theme)
    \begin{itemize}
        \item \textbf{Health Education \& Support:} Mentions that community health educators were a factor in health decisions
        \begin{itemize}
            \item Example: "A community health person came to my school and explained what was going on in the pandemic and the latest research on masking. So that's when we started masking."
        \end{itemize}
        \item \textbf{Support Incentives:} Mentions that a community-based organization provided an incentive or resource in some material form to support during the pandemic. For example, CBO (community-based organization) providing masks, hand sanitizer, gift cards, vaccination opportunities to community members.
        \begin{itemize}
            \item Example: "My local food pantry gave out masks which was super helpful when they were sold out online."
        \end{itemize}
        \item \textbf{Trust, rapport, \& relationships:} Mentions trust in community organization due to outreach, sharing personal stories, having open conversations, positive relationship-building
        \begin{itemize}
            \item Example: "It just made us feel like there was somewhere to turn to when everything was chaotic. The church gave us a place to talk about these things and feel safe."
        \end{itemize}
        \item \textbf{Reducing barriers:} Mentions a barrier, struggle, or challenge to implementing their health decisions or asserting agency in health decisions
        \begin{itemize}
            \item Example: "I just did not feel listened to by my doctor. And there was no alternative to what they were telling me. So"
        \end{itemize}
    \end{itemize}
    \item \textbf{Resources} (Theme)
     \begin{itemize}
        \item \textbf{Significant Impact Resources:} Mentions that impact resources like food, financial assistance, rent moratorium, student loan suspension, employment helped during COVID
        \begin{itemize}
            \item Example: "We would have been so lost without the food bank that restocked each week."
        \end{itemize}
        \item \textbf{Unmet community needs:} Mentions that a resource is needed
        \begin{itemize}
            \item Example: "We just never found the childcare we needed so that I could keep my job. I haven't worked since 2020."
        \end{itemize}
    \end{itemize}
    
    \item  \textbf{Vaccine Hesitancy} (Theme)
    \begin{itemize}
        \item \textbf{Did not vaccinate:} Mentions that the speaker did not choose to vaccinate
        \begin{itemize}
            \item Example: "I just couldn't get over how scary it was that my sister had this reaction to the vaccine. I know it could happen to me. So no I did not go through with the vaccine."
        \end{itemize}
        \item \textbf{Mistrust or Skepticism:} Mentions that the speaker has/had mistrust or skepticism of the vaccine
        \begin{itemize}
            \item Example: "There are people telling me this vaccine has a chip in it. I don't want a chip and I just have no way of knowing."
        \end{itemize}
    \end{itemize}
    
    \item \textbf{Personal COVID-19 Experience} (Theme)
    \begin{itemize}
        \item \textbf{Resilience, Connection, \& Hope:} Mentions agency, control, or feeling empowered during the pandemic period
        \begin{itemize}
            \item Example: "Helping at my church made me feel like I was making a difference even though the world was going crazy."
        \end{itemize}
        \item \textbf{Stress, Fear, \& Uncertainty:} Mentions stress, fear, or uncertainty during the pandemic period
        \begin{itemize}
            \item Example: "It was just anxiety all day every day thinking about my kid getting sick at school and bringing it back to her brother at home."
        \end{itemize}
    \end{itemize}

    \item \textbf{Future Visions \& Takeaways}
    \begin{itemize}
        \item \textbf{Conversation Reflections:} Mentions reflections on the conversation
        \begin{itemize}
            \item Example: "Talking about this has made me remember how hard that period was for our family."
        \end{itemize}
        \item \textbf{Post-pandemic future:} Mentions a future vision for their life or community after the pandemic
        \begin{itemize}
            \item Example: "I just can't wait for the schools to go back to normal and I hope we can all learn something from this."
        \end{itemize}
    \end{itemize}

\end{enumerate}

\section{RTFC Conversation Corpus Details}

\subsection{RTFC Conversation guide}

The conversation guide provided to facilitators included the following language, which guided facilitators to ask particular questions of participants.

\begin{itemize}
    \item Sharing Questions \& Lived Experiences
    \begin{itemize}
        \item As we shed the restrictions of the pandemic, it will be easy for us to lose sight of what we have learned about inequality in America and Boston and how our lived experiences have shaped that learning. Keeping this in mind, I’d like to invite you to think: “What’s your question about the future of Boston and your place in that future?”
        \item Thank you for sharing your questions with us. Now I’d like to invite you to think, what experience in your life got you to this question? 
    \end{itemize}
    \item Connecting Our Experiences
    \begin{itemize}
        \item Find someone whose question or experience resonates with your own life.  Then I want you to speak to that person and tell them why their question or experience resonated with you and share the story from your life that connects you with their experience.
    \end{itemize}
    \item Drawing Connections
    \begin{itemize}
        \item Let’s talk a little about what we are hearing. What are you hearing in people’s experiences? 
    \end{itemize}
    \item Wrap up
    \begin{itemize}
        \item Do you have any closing thoughts that you’d like to share or other general reflections? Do you have any questions for us?
    \end{itemize}
\end{itemize}

\subsection{RTFC Codebook}

Community partners, sensemakers, and stakeholders came together to develop the following codebook, which included the following themes of interest:

\begin{enumerate}
    \item \textbf{Government and Institutions} (Theme)
    \begin{itemize}
        \item \textbf{Expectations} References to the expectations and aspirations that the public has of elected officials, city government, and/or civic institutions. 
        \item \textbf{Processes:} References to processes through which the public interfaces with government, such as voting, community engagement, campaigning, electoral processes, and other decision-making processes. This may include feelings of exclusion, silencing, or neglect in public meetings; community dynamics within a public meeting; curiosity about electoral results; a lack of confidence in voting as a form of democratic participation. 
        \item \textbf{Accountability:}  Statements about the accountability of elected officials, city government, and/or civic institutions to the promises they make and the expectations they set for the public. This may include references to elected officials who "will tell you anything just to get your vote"; the city's failure to address pressing issues, like Mass and Cass; and general questions/doubts about how much the city listens to its residents and factors resident perspectives into decisions. E.g. "You said you were going to do this, and you have/but you haven't yet"
        
        \item \textbf{Institutional Resources:}  Statements about how people are having difficulty (or success) accessing services provided by government agencies and other institutions that improve one's quality of life. This could include mentions of services and resources like such as housing subsidies, senior services, mental health services, or municipal services like fixing potholes. This could also include statements about the difficulties people face in accessing these services or navigating institutions to get the services and resources they need.
        \item \textbf{ Community Resources:} Statements about how people are leveraging resources in their communities to fulfill their needs and improve one's quality of life. This could include mentions of community-based organizations that fulfill community needs; civic associations; or neighbors that provide support to other neighbors.
    \end{itemize}

    \item \textbf{Public Health} (Theme)
    \begin{itemize}
        \item \textbf{Mental Health:} Those who struggle with mental health; systemic issues of mental health; responses to those with mental health issues; resources and isntitutions that support mental health
        \item \textbf{Drugs and Drug Use Disorder:}  Addiction, systemic issues of drug use, responses to those with drug use disorders, the culture and environment around drug use
        \item \textbf{Trauma:}  Individual, community, generational traumas. The responses and resources intended to support healing from those traumas. Things that further cause traumas. 
        \item \textbf{Quality and Affordable Healthcare:} The accessibility, affordability, and quality of healthcare 
and other health services. 
        \item \textbf{Food Insecurity:}  Food access, quality of food accessible, food deserts, affordability of food, systems to support food accessibility.
        \item \textbf{COVID-19:} COVID-19, vaccines, mask, COVID tests, boosters, and the impacts of COVID-19 such as working from home, school closures, and jobs lost.
    \end{itemize}

    \item \textbf{Safety} (Theme)
    \begin{itemize}
        \item \textbf{Sense of Safety:} Refers to feeling unsafe within daily life routines at home, in one's neighborhood, and throughout the city.
        \item \textbf{Street Violence:}  Refers to situations like street fighting, assaults on the street, unintentional harm of bystanders, etc.
        \item \textbf{Gun Violence:}  Loss of family members due to a shooting, witnessing a shooting AND not limited to gang violence.
        \item \textbf{Policing:} Refers to being targeted by police (profiled) in certain areas and the lack of policing happening due to neighborhood location, race and/or ethnicity.
        \item \textbf{ Racialized Violence:}   Refers to verbal, emotional and physical assaults based on color of skin, race, ethnicity, language.
    \end{itemize}

    \item \textbf{Infrastructure} (Theme)
     \begin{itemize}
        \item \textbf{Climate Impacts:}  Climate change, impact of climate change on the community, actions to address climate change, fears around climate change.
        \item \textbf{Transportation:} Public transportation like the buses and trains, quality of transportation, affordability and accessibility of transportation, safety of public transit.
    \end{itemize}

       \item \textbf{Housing} (Theme)
    \begin{itemize}
        \item \textbf{Gentrification and displacement:} Displacement of lower income residents; physical transformation and change of the cultural character of the neighborhood.
        \item \textbf{Housing Instability:}  Difficulty paying rent, having frequent moves, living in overcrowded conditions, or doubling up with friends and relatives.
        \item \textbf{Homeownership:}  Challenges for owning a house; obstacles toward home ownership; expressing the with hope to be a home owner.
        \item \textbf{Housing quality:} the physical condition of a person's home as well as the quality of the social and physical environment in which the home is located
        \item \textbf{Housing affordability:} Cost of housing and how affordable that cost is to residents, regardless of tenure (tenant/owner), subsidy (e.g. workforce housing, public housing)
    \end{itemize}

    \item  \textbf{Community Life} (Theme)
    \begin{itemize}
        \item \textbf{Community Relationships:} Relationships between community members, across generations, and across communities. Quality and nature of those relationships.

        \item \textbf{Community Values:}  Values instilled throughout the community, values differences within and across communities.
    \end{itemize}
    
    \item \textbf{Education} (Theme)
    \begin{itemize}
        \item \textbf{Quality of Education:} Education that leads to empowerment as a process of strengthening individuals and communities to get more control over their own situations and environments; education systems that focus on the importance of quality learners, quality learning environment, quality content, quality processes, and quality outcomes
        \item \textbf{School Infrastructure:} Suitable spaces to learn; also spaces that have the infrastructure to address the COVID-19 public health emergency.
        \item \textbf{Life Skills:}The abilities (or the lack of) for adaptive and positive behaviour that enable individuals to deal effectively with the demands and challenges of everyday life in their communities and the world.
        \item \textbf{Youth Spaces:}Available and accessible physical and virtual spaces for activities especially offered to young people to advance their cognitive, emotional, social, and creative skills
        \item \textbf{Higher Education:} Post-secondary academic institutions, including colleges/universities/vocational schools, where individuals engage in advanced learning and research. Could be used to define relationships between students, teachers, administration.
    \end{itemize}

    \item \textbf{Economic Opportunity} (Theme)
    \begin{itemize}
        \item \textbf{Jobs:} References to a person's ability to provide for themselves and their families. Can include statements about working multiple jobs; working in a particular industry; facing unemployment; job satisfaction; difficulties in finding a job; observations about the job market; discrimination within a job or during a job search; efforts to attain more training or education in order to improve one's job prospects 

        \item \textbf{Economic Assistance:} References to one's ability to access economic supports that enable wealth-building, financial stability, and/or economic growth. This can include statements about individuals (such as one's ability to access home loans) and small businesses (such as a business's ability to access lines of credit).

        \item \textbf{Income:} Explicit references to income/wages and wealth. This can include discussions about: one's personal income; satisfaction with their income; in/ability to increase their income; in/ability to build wealth; income inequality; the income/wage levels to be able to afford the cost of living in Boston. 

        \item \textbf{Affordable Childcare:} References to one's ability to afford childcare. This is included in Economic Opportunity because childcare affects one's ability to maintain stable employment.

        \item \textbf{Financial Literacy:} References to people's level of financial literacy, from everyday money management, to processes for applying loans and credit. This can also refer to people's general lack of financial literacy.
    \end{itemize}

    \item \textbf{Inequality} (Theme)
    \begin{itemize}
        \item \textbf{Race:} Defined as lack of jobs, services, goods, based on skin color, ethnicity, language.

        \item \textbf{Class:} Refers to socioeconomic status, education, and types of disparities, including neighbors re-entering society.

        \item \textbf{Gender:} Discrimination based on (anatomy) female, male.

        \item \textbf{Sexual Orientation:} Refers to sexual identity and preference.

        \item \textbf{Ability:}  Refers to disabilities, physical and intellectual.

        \item \textbf{Immigration Status:} Foreign born, regardless of documentation - this example speaks more to being an immigrant in which English is the second language, which is the barrier of an immigrant.

    \end{itemize}

\end{enumerate}
 
\twocolumn

\subsection{Prompts to GPT-4 and Llama}

We provided an instruction, list of labels and definitions in the codebook in JSON format for each corpus' codebook. We requested output in JSON format as follows:

\lstset{
    basicstyle=\ttfamily\footnotesize, 
    breaklines=true,                    
    breakatwhitespace=false,             
    columns=fullflexible,                
    frame=single,                         
    keepspaces=true                       
}

\begin{lstlisting}
Your job is to provide a comprehensive set of thematic labels for the given quote.
You are given 7 [9] thematic tagging questions, subthemes, and general descriptions\\ in the JSON below.
Choose ONLY from the tags provided here.

"annotation_schemes": [<list of top-level labels and sublabels, in JSON format, with definitions>]

For the given conversation quote, return all subtags that apply in a single JSON array in this format, (or return ``None of the above::I confirm that none of the themes apply" if none apply or if the statement is too ambiguous to determine):

Expected format:
```json
[
  {{
    "highlight_id": highlight_id,
    "tags": ["External Motivations::Employers", "Intrinsic Motivations::Not wanting to get the virus"]
  }}
]
```
Please share this output format with no any additional characters, annotations, line breaks, or comments.

highlight_id: [id for excerpt]
Conversation quote: [quote] 
JSON:


\end{lstlisting}
Additional details on prompts and code are available at \href{https://github.com/schropes/ai\_sensemaking}{https://github.com/schropes/ai\_sensemaking}.
The models were prompted at a temperature of 0. 
Overall, responses were very well-formed according to this prompt, both for Llama and GPT-4. One single label that was not in our codebook was hallucinated 3 times across the generation process for the NYC corpus: \textit{External Motivations: Government}. Qualitatively, we note with interest that this theme did come up often in the conversation quotes, and therefore could be seen as a thematically relevant code despite the violation of instructions needed to produce it. 
Two labels were hallucinated in the RTFC codebook: "Community Life: Community Resources" and "Housing: Housing Stability." Both of these were also plausible given the conversation content, and were hallucinated just once.
All hallucinated labels were removed to ensure fidelity to the original codebook. In one instance, GPT generated "None of the above" in addition to another valid label. For this case, we removed the nonsensical "None of the above" label from the suggested labels. 
In one case, GPT-4 failed to produce a label of the requested format for the RTFC data. We converted this suggested label into "None of the above" for consistency with the rest of the corpus.

\section{Crowdworker recruitment}
Crowdworkers were recruited from Prolific, and had the following characteristics: located in the US or UK, with English as a first language, with 95-100 percent approval ratings, who had attended some or more college to participate in our annotation experiment. We paid the recommended rate on Prolific of \$12/hour, 1.6x the US minimum, and adjusted upward to \$12/hour if our initial estimated time was not sufficient to pay workers this amount.

\subsection{Annotator "understanding" questions}

From the round of IRR validation done before the main round of experiments, we found several examples of very high agreement quote-label pairs. We selected four of them to act as test questions for annotators. Two examples of these high-confidence test questions for the NYC: corpus are listed here.

\begin{itemize}
    \item High agreement example of \textit{Family \& Friends} label: "Right, but you know what? I encouraged my sons and my daughter to vaccinate their kids because you don't know that COVID is new. You don't know how it's going to affect them and the children, my grandkids."
    \item High agreement example of \textit{Support Incentives:} "I think when I made my decision it came in handy, because I think it was a month before my daughter was starting school and they were giving you \$100 for the vaccine. And honestly, that came at a good time. Why? Because I said, `Okay, they gave me \$100. They gave her \$100.' And I said, `Oh, this is good because now I could get you this and that before school starts.' So that was pretty good. I mean, that was nice." 
\end{itemize}

\subsection{Instructions to annotators}

Following a consent page, annotators received the following instructions:

\textit{"First, we need to explain the task we are asking you to complete. We will ask you to read a quotation from a conversation. The conversation is about <resources and challenges during the COVID-19 pandemic in the United States. Conversations were hosted to better understand resources that helped during the pandemic, challenges to access, and motivations for making health-related decisions during the pandemic>.}

\textit{We are asking you to identify <7> phenomena in this annotation. Please read about each type before moving forward by going here: [link to Codebook Training Document]. We will check that you have spent at least 3 minutes reading this document before advancing to the task.}

\textit{There are 15-22 annotations in the task. Review your answer for each question before proceeding."}

Then, annotators were given the following choices: \textit{"I agree to read carefully and spend enough time on each annotation"} or \textit{"I do not wish to partake"}. At the bottom, text read \textit{"At the top of the next page, there will be a quote. To annotate, select the check boxes that apply."} In the LLM assistance conditions, this text read: \textit{"At the top of the next page, there will be a quote, followed by a list of suggested labels for the quote. Please read the quote and the list suggested labels, which you may use to assist your annotation. To annotate, select the check boxes that apply."} If they had selected the option to proceed with the study, stimuli to annotate were next presented. After 20 annotations + 2 test questions were presented, participants were given the option to provide demographic information for research purposes.

The crowdworker study was reviewed by the MIT IRB review board, and determined exempt.

\section{AI Assistance}
In this work, the authors used some AI tools to find related works, including Elicit and ScholarQA. Citations were followed and checked. We also used Github Copilot and Cursor for coding assistance, and code was reviewed for errors.

\end{document}